# An Integrated Design and Verification Methodology for Reconfigurable Multimedia Systems

*M. Borgatti, A. Capello, U. Rossi*
*STMicroelectronics, Agrate, Italy*
michele.borgatti@st.com
andrea.capello@st.com
umberto.rossi@st.com

*J.-L. Lambert, I. Moussa*
*TNI-Valysosys, France*
jean-luc.lambert@tni-valiosys.com
imed.moussa@tni-valiosys.com

*F. Fummi, G. Pravadelli*
*University of Verona, Italy*
franco.fummi@univr.it
pravadelli@univr.it

**Abstract**

*Recently a lot of multimedia applications are emerging on portable appliances. They require both the flexibility of upgradeable devices (traditionally software based) and a powerful computing engine (typically hardware). In this context, programmable HW and dynamic reconfiguration allow novel approaches to the migration of algorithms from SW to HW. Thus, in the frame of the Symbad project, we propose an industrial design flow for reconfigurable SoC's. The goal of Symbad consists of developing a system level design platform for hardware and software SoC systems including formal and semi-formal verification techniques.*

## 1. Introduction and motivations

The recent introduction of embedded programmable logic allows application-specific integrated circuit (ASIC) and application-specific standard product (ASSP) vendors to broaden the versatility of their products. Dynamic HW reconfigurability is becoming a popular concept [1][2][3]. Different technologies can implement this concept, but the so called "HW virtualization" based on field-programmable gate arrays (FPGA's) is the one where the practical tradeoff among performance, size, power consumption and costs can be achieved for a larger number of final applications and not only prototypes.

Reconfigurable FPGA's are particularly suited for multimedia applications on portable appliances. In fact, tomorrow's multimedia applications will require both the flexibility of upgradeable devices, traditionally software-based, and a powerful computing engine typically embodied in hardware. Reconfigurable hardware (RH), can meet both these requirements, being the performance of a specific task executed in HW much faster than the performance of the same task executed in SW. Multimedia application domain is therefore a very good target for RH architectures.

Due to complexity of reconfigurable architecture, the design and verification phases cannot be independent processes. Thus, the goal of the Symbad project is to develop a system level design framework for hardware and software SoC systems including formal verification techniques and automatic test pattern generation (ATPG). Formal verification is applied to specific problems related to reconfigurability, while ATPG is used to detect design errors in the early phase of design flow.

This paper describes a user scenario that motivates the introduction of reconfigurable hardware into industrial applications together with a vision on the platform, called Vista, that should be built to support reconfigurable computing. This platform and its verification techniques will be assessed on the design of a reconfigurable SoC targeted to multimedia applications. Moreover, the paper emphasizes the use of formal and semi-formal techniques during the verification process.

## 2. Configurable platform architecture

The proposed methodology is assessed with the design of a reconfigurable image processing system where the combinatorial complexity of reconfiguration makes simulation, testing and verification so long, with existing techniques, to make it unpractical for the usage in the field. The impact of Symbad framework is important on the productivity of design teams, optimization and reliability of systems, and the development of SoC products or embedded systems. In the frame of Symbad we started with a stable design flow based on classical approach, including:

I. Concept validation performed at the "C" level.
II. Modeling by a number of tasks, still in "C", where abstract communication is introduced.
III. Profiling of the various tasks based on the application execution.
IV. Mapping on HW and SW resources.
V. Mapping parts of HW onto FPGA.

The actions in the list constitute the architecture exploration process, where a single configuration must be graded according to performance, silicon usage, power consumption. This process includes a number of iterations through II-III-IV steps to find the best product trade-off.

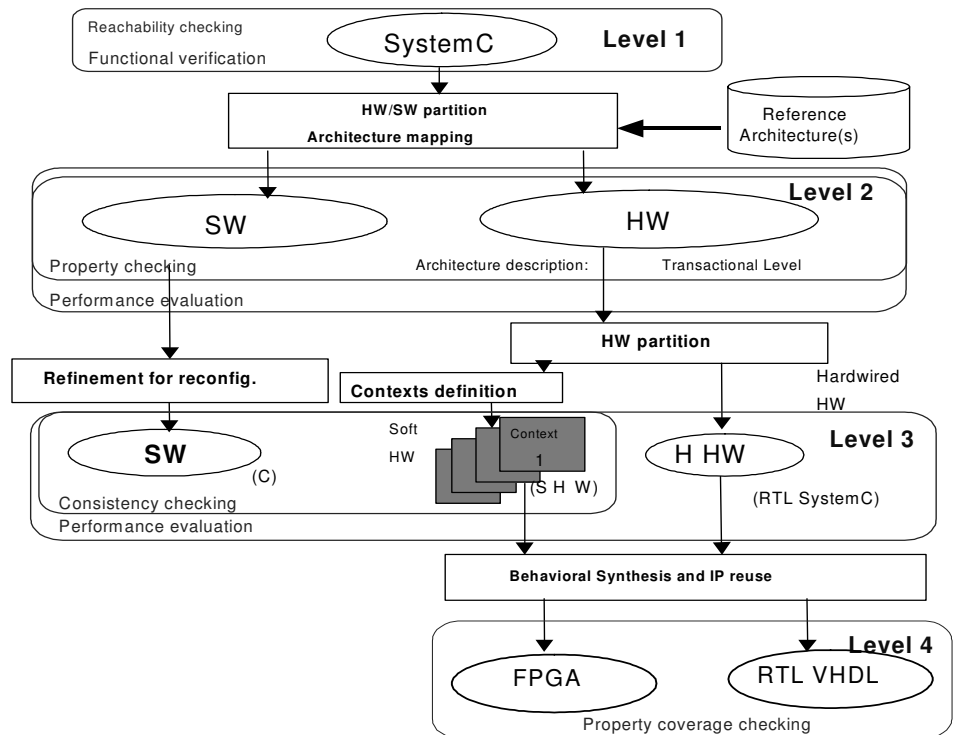

Figure 1: **Proposed design and verification flow (Symbad).**

Validation is performed once for the initial concept description. Then, the verification of each iteration step has to be postponed at the IV stage, where HW is described in RTL and simulated at cycle level, and SW is executed on an instruction set simulator (ISS) of the general purpose processor. Then the final system verification is performed after mapping HW onto FPGA. This approach was considered acceptable for prototyping the proposed silicon technology, but is definitely unsatisfactory to be deployed in production, being the verification of each architectural exploration step of the order of tens of hours. A remedy to simulation slowness can be found in HW/SW co-emulation/simulation, but this solution is still too expensive and requires too many specialists to be economically feasible.

Rather than building a brand new system design flow, it is desirable to add useful features at I-II-III-IV stages by providing better analysis capabilities and improving predictability of the whole design flow. New technologies are needed for this flow in order to enhance this approach. These new technologies are:
- simulation at transactional level;
- formal/semi-formal verification.

Transactional level (TL) modeling is proposed as a way to minimize the amount of events and information processed during simulation to dramatically speed up the validation time. In TL the communication is completely separated from computation, and the focus is on the data rather than on the way the transfer is executed. In the traditional previously described flow, transactional simulation is only used in phase I, II and III at "C level". We propose to extend its use to other phases by introducing it:
- At stage IV, by doing the simulation of a SystemC model of the HW/SW mapping in order to do performance evaluation. The speed of simulation being guaranteed by the application software running on the host machine (without any need for ISS use).
- At stage V, by adding to the model a modeling of the FPGA reconfiguration. Here again the objective is to do performance analysis taking into account the downloading of bit streams through the bus.

This transactional level simulation is run with the help of libraries and extensions of Vista tool [4].

On the other hand, formal and semi-formal verification can be profitably applied at several stages of the above approach as described in Section 3. Four approaches are exploited in a cascade fashion to address different verification problems at different design levels: ATPG to quickly remove easy-to-detect design errors on the behavioral description, linear programming verification to verify real-time properties when timing information is introduced, abstract interpretation to check reconfiguration consistency after FPGA mapping, and model checking to verify the correctness of the final RTL description.

## 3. Design flow methodology

Transactional level simulation and formal/semi-formal verification can be included in the traditional design flow described in Section 2. To accomplish the goal we propose a novel methodology for designing and verifying reconfigurable SoC's. It is divided in four levels as shown in Figure 1.

### 3.1. System level specification: level 1

In level 1, the flow begins with a purely functional description of the system, there the system can be simulated with the help of the standard SystemC simulator. This permits to check that basic functionalities are actually realized by the system. At that level, one does not know which SystemC entities will be mapped onto hardware, software or reconfigurable hardware.

At this level, functional verification is applied by using a SystemC-based ATPG (Laerte++ [5]) to estimate the coverage of test benches. The test pattern generator exploits both simulation-based techniques, (e.g., genetic algorithms) and formal-based ones (e.g., SAT-solvers). Coverage measures are based on standard metrics (statement, condition and branch coverage) and on the more accurate bit-coverage metric exploiting high-level faults [6]. This information is used to quickly identify potential design errors.

Moreover, a new technology based on linear programming verification (LPV [7]) is used for proving deadlock freeness. The SystemC model is translated in an abstract model where communication and synchronization characteristics remains un-abstracted. Then deadlock situations are checked formally, each deadlock situation being translated in an unreachability property. These properties can be automatically generated. Note that only deadlock situation captured as unreachability property can be check by this mean, LPV being only able to deal with reachability problems.

### 3.2. Architecture mapping: level 2

At level 2, the description obtained is mapped onto an architecture. This architecture mapping consists in deciding HW/SW partitioning and in providing the HW with a communication architecture (busses, point to point communication, shared variables, etc). During this level, simulation is used intensively for evaluating the different possible architectures. The goal is to get the best compromise between, for example, power consumption, bus loading and memory accesses.

This level is a good target for formal verification issues. It is also the level where the system performance analysis can be applied by using the Vista tool. This later can be used as it provides the user with libraries for representing SystemC models of busses, peripherals and memory elements. But this second phase does not take into account the partition between pure HW and reconfigurable HW (often called soft hardware).

In that phase, LPV is used to prove real-time properties like timing deadline achievement and FIFO channel dimensioning.

### 3.3. Architecture refinement and reconfiguration: level 3

Reconfigurability issues appear at the third level. Here the HW is separated in pure HW and reconfigurable HW. It is then necessary to refine the previous analyses by simulating a model of the system where the bit streams download, due to reconfigurations, is part of the bus loading. To do this, it is strictly necessary to introduce the reconfigurability orders in the SW, and to provide libraries for FPGA reconfiguration modeling. Finally, in order to evaluate timings, the SW is annotated.

The Vista tool is used for evaluating the impact of the reconfigurable hardware characteristics on the performances of the system. The characteristics of the reconfigurable hardware consist in a set of FPGA configurations which can be changed by the software at run-time. Each configuration contains a fixed set of computing resources (in the Symbad case study: some HW modules implementing algorithms and registers). The partition of algorithms and registers among the different configurations is an important architectural aspect which must be thoroughly tuned for obtaining optimal performances. Unfortunately, the modification of the software by introduction of reconfiguration instructions cannot be done in an automatic manner. The reason concerns the optimization of the system by reducing the number of reconfigurations. Indeed, downloading bit streams is costly in terms of bus loading and it is rather tricky to ensure automatically a good reduction of them.

Another tool, called SymbC, is provided by the Symbad project for formally verifying that the modified SW satisfies the following fundamental consistency property: *"each time the software requires a hardware resource of the reconfigurable part, this resource is actually available"*.

Note that this property is only SW dependent, since in the frame of Symbad, the software is lonely responsible for initiating an FPGA reconfiguration. Symbc takes at the input:
- The application C code containing FPGA reconfiguration instructions and resource calls C code.
- A configuration information containing:
  - The name and signature of the reconfiguration procedure.

- The name of the functions that are implemented in the FPGA (and that can be absent from it).
- The FPGA configuration characteristics (i.e., which function is present in which configuration) and provides at its output a certificate of consistency (proving formally that any functions is only invoked when it is present in the FPGA) or a counter-example showing a problem.

### 3.4. RTL generation: level 4

At level 4, the RTL code is produced. Depending on the architecture chosen at level 2, some properties are defined to formally check the correctness of the HW/SW interface. Model checking and SAT solving are used at this level [8][9]. However, proven properties cannot completely assure the correctness of the design implementation, since some behaviors may have been not considered. Thus, how many properties should the verification engineer define to completely check the implementation? Few works, based on symbolic methods, are related to the properties incompleteness topic [10][11][12], but their applicability is limited by the state explosion problem. To solve the problem, we have developed a tool, called property coverage checker (*PCC*), that evaluate the completeness of properties by mixing functional and formal verification [13].

The designer uses a model checker to prove properties on the RTL model. Either a proof certificate or a counter example is expected for each property. The design needs to be revised each time a property failure is obtained. When all properties have been proved, the PCC is used. If it shows that not enough properties have been used, again, the designer will have to extend the set of properties and check the new ones. The cycle continues until no more refinement is possible.

### 4. Case study

The proposed design and verification methodology has been applied to a *face recognition* system by mapping the application to a reconfigurable platform. The nature of the reconfigurable platform allows specifications of the system to translate to the target implementation, leaving flexibility to possibly implement other applications of the same family. The target application consists of recognition of a face previously acquired by a low-resolution CMOS camera. The recognition phase is performed comparing the unknown face to a database of twenty different faces under multiple poses. Applications are low cost smart toys, advanced human-machine interfaces and color CMOS camera processors.

The reference model of the complete system functionality is a collection of programs written in C. A first implementation of the face recognition system was built upon a reconfigurable platform based on embedded FPGA and an extensible 32-bit microprocessor. This implementation hase been obtained by following a top-down methodology without specific focus on reconfigurable systems. The design flow was based on a "static" approach where all HW resources being implemented were assumed to be simultaneously available in the system. Moreover, FPGA definition and consistency check was done manually. This resulted to be a difficult and error prone process.

Out of the same reference model a new design and implementation process has been done following the proposed methodology. This includes transaction-level modeling and architecture exploration as well as formal checks oriented at the consistency of reconfigurable systems. As seen in Section 3 the methodology is articulated into four different refinements of the system description.

### 4.1. Design exploration

The level 1 description is a pure functional un-timed point-to-point communication model written in SystemC 2.0. Referring to Figure **2**, CAMERA is the abstract representation of a CMOS camera device, DATABASE is an abstract representation of a nonvolatile memory system that will be eventually implemented to a flash memory device. At this level of abstraction simulation is performed at transactional level and its results can be matched against the C reference model. Match of results consists of trace files comparison as the TL model captures data consistently to the reference one.

The complete simulation of the system TL model took less than 15 seconds when executed on a Sun U80 dual-processor workstation running Solaris 2.8 OS. The functionality was fully verified against the reference model and the debug was eased by the untimed nature of the model. This step of the flow was completed in a couple of weeks starting from the availability of the reference model.

At level 2, architectural exploration begins. Within the system modeling and simulation environment (Vista) the designer was supported in automating the partitioning of the level 1 system description into HW and SW. SW modules have been collapsed to a single large SW task . This task models the SW partition of the system being executed into a CPU model (ARM7TDMI in the actual design) and corresponds to a simple cyclostatic scheduling for the 10 original SystemC modules. No further modeling of operating system functionalities can be done at this level of system description.

This HW/SW partition is based on designer's knowledge about the heaviest computational tasks. This ranking of the most demanding tasks is done by execution profiling of the UT code developed at level 1. Therefore accurate profiling is of key relevance to estimate performance of the architecture under investigation.

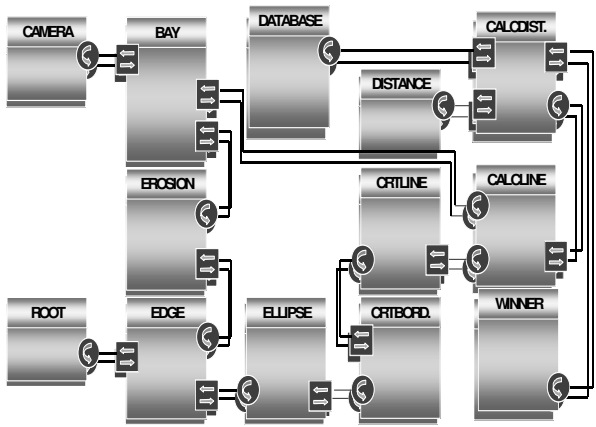

**Figure 2:** Level 1 face recognition system

Timing information is the most important system figure at level 2. Cycle accurate timing of SW can be automatically extracted by Vista based on a library of model(s) of available processor(s). Annotation into SystemC models of SW part is fully automated. Annotation refers to the execution time of the embedded SW that will eventually run on the target CPU. This means that simulation uses SystemC code modeling the embedded SW for the purpose of timing estimation only. Therefore it is possible to add code into the systemC model (for instance to ease debug) without affecting the timing figures. This is the case of `printfs` or file-system calls that are executed but skipped for timing annotation (unless they belong to the original code).

Also, suitable TL timing information must be annotated into SystemC models of HW parts. Reasonable assumptions on HW timing rely on designer's experience on performance of HW logic and coprocessors into the target technology. Annotation is manual for HW models.

In addition to the feature of timing annotation, another automation that is provided to help the architectural exploration phase is related to structural modification of the architecture under investigation. There are two main transformations that are required:
1. Transforming the UT model to the TL timed by adding one or more connections (buses, X-bars, etc).
2. Incrementally modify the TL timed model to move tasks between the HW and SW partitions.

Transformation 1 is made up of the following elementary operations:
- Grouping the first candidate SW into a single task featuring the union of all point-to-point connections.
- Instantiating the SW task into the selected CPU model featuring a single bus interface.
- Instantiating connection resources.
- Connecting the CPU model and all HW parts to the connection resources.

Transformation 2 can be divided into two basilar operation:
- Moving one module from HW to SW side.
- Moving one module from SW to HW side.

Each transformation foresees to build a new wrapper for the SW side and, eventually, to add or remove a connection to the connecting resource. Profiling and annotation have to be repeated for the new SW task, but it's an automated feature of the system modeling and simulation environment (Vista). For the hardware side, timing annotation must be done only in the case of modules moved from the SW to HW.

The TL model of the partitioned system is able to produce a simulation speed closed to 200kHz when executing on a Sun U80 dual-processor workstation, running Solaris 2.8 OS. Functionality has been fully verified matching the results against the level 1 ones. One week has been the time cost to perform the architectural exploration of the system, including the profiling step, annotations of both HW and SW side and collecting statistics of the final architecture.

Level 3 of the methodology flow is the heart of the reconfigurable platform. Here the dynamic reconfigurable device (FPGA) is instantiated into the design and some of the HW modules, obtained from the previous HW/SW partitioning, are carried inside the FPGA.

Moving functionality from pure HW to FPGA, or viceversa, is not a demanding task. Operation steps to perform the mapping are described below:
- Instantiating the FPGA SystemC model into the design and connect it to the connecting resource (bus).
- Disconnecting the HW modules from bus and connecting to the FPGA, defining the appropriate contexts.
- Inserting the FPGA's reconfiguration calls and the functional calls to mapped resources into the SW.

For the target architecture under investigation it has been quite reasonable that modules DISTANCE and ROOT be mapped both into the FPGA. They have been spitted into two different contexts, named config1 and config2. Manual instrumentation of the SW code has been performed, that is a specific configuration is loaded into the FPGA before the functions that belongs to it are called. The FPGA context switch becomes relevant in evaluating the system performance, so the same analysis performed at level 2 is to be applied to confirm the effectiveness of the designer's choice about the FPGA resource mapping.

The simulation speed of this level of the methodology flow is closed to 30kHz when executing on a Sun U80 dual-processor workstation, running Solaris 2.8 OS. Functionality has been fully verified matching the results against the level 2 ones. Less than one week was required to perform the mapping of the HW modules into the FPGA, the integrity check of the software and to collect performance reports for the architecture under analysis.

Level 4 represents the final mapping of the chosen architecture. The complete task of mapping the SystemC to RTL, a.k.a behavioral synthesis, is much farther the purpose of Vista. In our test case we can easily support a few pre-defined IP's, mainly concerning the CPU, the connection resource (AMBA bus), the FPGA and the memory. Automated interface synthesis is part of the foreseeable options, and also checkers for those interfaces could be automatically generated. For the current design, interface synthesis between SW side and HW parts, that is the construction of dedicated wrappers to convert RTL SystemC protocol, used by HW modules, to transactional level, used by the connection resource, was manually performed for each HW module. One week has been spent to build the interfaces, time that could be significantly reduced by the automation of the phase.

### 4.2. Design verification

The SystemC description realized at level 1 has been verified first by using Laerte++. The memory inspection capability of Laerte++ allows us to quickly identify and remove design errors related to incorrect memory initialization. These errors reflected on a less precise images matching. On the other hand, the application of LPV allowed efficient hunt of deadlock conditions.

At level two, the HW/SW partitioning and the introduction of an AMBA bus required a new verification phase focused on timing issues. ATPG is not suited to detect timing errors, thus, LPV has been used to prove real-time properties like timing deadline achievement and FIFO channel dimensioning.

After reconfigurable device instantiation, the full integrity of the design has been tested by application of SymbC. This assured that for any path of the application's control flow the FPGA was loaded with the necessary functions.

Finally, model checking has been applied at level 4. Formal properties related to the correct implementation of critical RTL modules have been defined. The adoption of PCC allowed us to identify property missing in the initial verification plan that none of previous verification phases have revealed.

### 5. Conclusion

The characteristics of a powerful design and verification flow, featuring semi-formal and formal techniques, have been reported together with a test case to benchmark the effectiveness of the novel approach. A vision was presented on the architectural challenges and the required programming environment for reconfigurable platform. Moreover, a verification strategy is proposed that efficiently exploits different techniques at different design levels.

### 6. References

[1] D.Panigrahi, C.N.Taylor, S.Dey, "*A Hardware/Software Reconfigurable Architecture for Adaptive Wireless Image Communication*", Proc. ASP-DAC, pp.553-560, 2002.

[2] Y.Li, T.Callahan, E.Darnell, R.Harr, U.Kurkure, J.Stockwood. "*Hardware-software co-design of embedded reconfigurable architectures*", Proc. DAC, pp. 507-512, 2000.

[3] D.Verkest; D.Desmet; P.Avasare; P.Coene; S.Decneut; F.Hendrickx; T.Marescaux; J.Y.Mignolet; R.Pasko; P.Schaumont: "*Design of a Secure, Intelligent, and Reconfigu-rable Web Cam Using a C Based System Design Flow*", Proc. Asilomar Conference on Signals Systems & Computers, pp. 463-467, 2001.

[4] I. Moussa, T. Grellier, and G. Nguyen, "*Exploring SW Performance using SoC Transaction-level Modelling*", Proc. DATE, pp. 120-125, 2003

[5] A.Fin, F.Fummi, "*Laerte++: an Object Oriented High-Level TPG for SystemC Designs*", Proc. FDL, 2003.

[6] F. Ferrandi, F. Fummi, and D. Sciuto, "*Implicit Test Generation for Behavioral VHDL Models*", Proc. ITC, pp. 436-441, 1998.

[7] S. Dellacherie, S. Devulder, and J-L. Lambert, "*Software verification based on linear programming*". LNCS, Vol. 1709, pp.1147-1165, 1999

[8] K.L.McMillan, "*Symbolic Model Checking*", Academic Press, Norwell, MA, 1993.

[9] I.Beer, S.Ben-David, C.Eisner, A.Landver, "*Rulebase, an Industry-Oriented Formal Verification Tool*", Proc. DAC, pp. 655-660, 1996.

[10] Y.Hoskote, T.Kam, P.H.Ho, X.Zao, "*Coverage Extimation for Symbolic Model Checking*", Proc. DAC, pp. 300-305, 1999.

[11] S.Katz, O.Grumberg, D.Geist, "*Have I Written Enough Properties? – A Method of Comparison between Specification and Implementation*", Proc. CHARME, pp. 280-297, 1999.

[12] H.Chockler, O.Kupferman, R.P.Kurshan, M.Y. Vardi, "*A Practical Approach to Coverage in Model Checking*", Proc. CAV, pp. 66-78, 2001.

[13] A.Fedeli, F.Fummi, G.Pravadelli, U.Rossi, F.Toto, "*On the Use of a High-level Fault Model to Check Properties Incompleteness*", Proc. MEMOCODE, pp.145-152, 2003